\documentclass[aps,pra,twocolumn,superscriptaddress,floatfix]{revtex4-1}
\usepackage{blindtext,amsmath}
\usepackage{graphicx}
\usepackage{float}
\usepackage[export]{adjustbox}
\newcommand{\ophi}{\hat{\phi}(x)}
\newcommand{\odphi}{\hat{\phi}^\dagger(x)}
\newcommand{\gbb}{g_\mathrm{BB}}
\newcommand{\gib}{g_\mathrm{IB}}
\newcommand{\mred}{m_\mathrm{r}}
\newcommand{\oxi}{\hat{\xi}(x)}
\newcommand{\odxi}{\hat{\xi}^\dagger(x)}

\usepackage{color}

\begin{document}
\title{Strong-coupling Bose polarons in 1D: Condensate deformation and modified Bogoliubov phonons}
\author{J.~Jager}
\affiliation{Department of Mathematics, Imperial College London, London SW7 2AZ, United Kingdom}
\author{R.~Barnett}
\affiliation{Department of Mathematics, Imperial College London, London SW7 2AZ, United Kingdom}
\author{M.~Will}
\affiliation{Department of Physics and Research Center OPTIMAS, University of Kaiserslautern, 67663 Kaiserslautern, Germany}

\author{M.~Fleischhauer}
\affiliation{Department of Physics and Research Center OPTIMAS, University of Kaiserslautern, 67663 Kaiserslautern, Germany}

\begin{abstract}
We discuss the interaction of a quantum impurity with a 
one-dimensional 
degenerate Bose gas forming a Bose-polaron. In three spatial dimensions 
the quasiparticle is typically well described by the extended Fr\"ohlich model, in full analogy with the solid-state counterpart. This description, which assumes an
undepleted condensate, fails however in 1D, where the backaction of the impurity on the condensate leads to a self-bound mean-field polaron for arbitrarily weak impurity-boson interactions. We present a model that takes into account this backaction and describes the impurity-condensate interaction as coupling to phonon-like excitations of a deformed condensate. 
A comparison of polaron energies and masses to diffusion quantum Monte-Carlo simulations shows very good agreement already on the level of analytical mean-field solutions and is further improved when taking into account quantum fluctuations.
\end{abstract}

\maketitle

\section{Introduction}

 The 
 polaron, introduced by Landau and Pekar \cite{Landau1933,Pekar1946} to describe
the interaction of an electron with lattice vibrations 
in a solid, is a paradigmatic model of quasiparticle formation in condensed matter physics. A hallmark feature of the quasiparticle is mass enhancement: the  electron becomes dressed by a cloud of  phonons
which in turn affects its dynamical properties. 
The polaron concept has wide applications across condensed matter physics ranging from  charge transport in organic semiconductors to high-$T_c$ superconductivity \cite{Mott1993,Alexandrov2010}.

 More recently, neutral atoms immersed in quantum gases have attracted much attention since they are experimentally accessible
platforms for studying polaron physics with high precision and in novel regimes. For example, the impurity-bath interaction
can be tuned from weak to strong coupling
employing Feshbach resonances \cite{Chin2010}. 
In such systems, the impurity atom is immersed in a superfluid and a polaron is formed by its interaction with the collective excitations of the superfluid. 
The Fermi-polaron, i.e.\  an impurity in a degenerate Fermi gas has been studied in a number of experiments  \cite{Schirotzek2009,Zhang2012,Kohstall2012,Koschorreck2012,Scazza2017,Cetina2015,Cetina2016,Parish2016,MassignanRPP2014}.  In contrast, only a few experiments on Bose polarons exist \cite{Catani2012,Jorgensen,Hu,Yan2019}.  Due to the  compressibility of a Bose gas, a large number of excitations can be generated, and interactions within the Bose gas are important.

 Theoretical works addressing the Bose polaron 
most often describe the interaction with the impurity as a coupling to Bogoliubov phonons of a \emph{uniform} superfluid
\cite{Rath2013,Astrakharchik,Tempere,Casteels2011,Grusdt2017b,Shchadilova2016}. The resulting (extended) Fr\"ohlich Hamiltonian 
is formally identical to the one used in solid-state systems \cite{Frohlich1954}, amended with two-phonon scattering terms. Efficient  approaches for its solution beyond the perturbative regime
have been developed in the past, including variational \cite{Li-PRA2014,Levinsen-PRL2015,Shchadilova2016}, field-theoretical \cite{Rath2013,Casteels-PRA2014,Christensen-PRL2015,Ichmoukhamedov2019}, renormalization group (RG)  \cite{Grusdt2015,Grusdt2017b} and
open-system approaches \cite{Lampo2018}, as well as Quantum Monte-Carlo simulations \cite{Ardila2015,Parisi2017a,Grusdt2017b}.
However, as well known from the example of electrons in superfluid Helium, a strongly interacting impurity can also distort the superfluid itself \cite{Hernandez1991}. This deformation creates a potential for the impurity which can lead to a self-bound state.
In 3D, the normalized impurity-Bose interaction 
has to exceed a critical value for this, given by the inverse gas parameter 
\cite{Cucchietti2006,Bruderer-EPL2008,Blinova2013}. 
  Since for typical condensates the gas parameter is very small, the extended Fr\"ohlich model remains adequate. 

 The situation is different in 1D, which was experimentally realized in \cite{Catani2012}. Here  an arbitrarily weak deformation of the condensate leads to a self-localized impurity 
 \cite{Bruderer-EPL2008}.  This restricts the accuracy of the Fr\"ohlich model to the perturbative regime. In fact a comparison between exact diffusion Monte-Carlo (DMC) simulations of the full  model
 with RG solutions of the extended Fr\"ohlich model in
 \cite{Grusdt2017b} shows that this model is only 
  accurate for weak interactions and breaks down completely for attractive interactions at intermediate interactions. 

 In this Letter, we follow a different approach, and expand the Bose quantum field about the exact mean-field solution in the presence of the mobile impurity in the Lee-Low-Pines (LLP) frame \cite{Lee1953}. 
 Such a treatment incorporates the backaction of the impurity already at the mean-field level 
 as in  \cite{Cucchietti2006,Bruderer-EPL2008,Blinova2013}, but keeps the entanglement between impurity and BEC by working in the LLP frame. 
  Quantum effects are then taken into account by the coupling to phonon-like excitations of the deformed superfluid. 
  Motivated by experiments  \cite{Catani2012} and the availability of semi-analytic mean-field solutions, we here consider a 1D quasi condensate with weak to moderate boson-boson interactions. While the experiments are performed in a harmonic trap, we  assume periodic boundary conditions when introducing  phonons. Stricly speaking  there is no BEC in a homogeneous 1D system and also the quasi-particle concept is believed to break down \cite{Kantian-PRL2014,Lausch2018} due to a diverging number of low energy excitations emitted by the impurity. Thus special care must be taken when calculating quantum effects. 
 We derive the effective Hamiltonian for the deformed phonons and solve them in  Bogoliubov approximation. Our treatment carries over naturally to higher dimensional systems with the only difference that the mean-field solutions have to be obtained numerically.
 Other treatments of the 1D polaron based on a factorization of the $N$-particle wavefunction
 in the LLP frame exist that take the deformation of the condensate into account 
 \cite{Hakim1997,Volosniev2017,Mistakidis2019,Panochko2019}. The scope of extending them  to incorporate quantum fluctuations is limited, however.
We note that the standard arguments
to define the polaron mass, applicable for Fr\"ohlich-type models, give non-sensical results here and require a careful reconsideration.
 We derive analytical expressions for the mean-field polaron wavefunction, from which we
 reproduce previous approximations for the polaronic mass and energy. We then calculate quantum corrections by solving the Bogoliubov deGenne equations
 in a self-consistent approach.
 Our results are benchmarked against recent DMC results \cite{Grusdt2017b}. We find very good agreement in all regimes for repulsive interactions underpinning the hypothesis that expanding about the non-uniform condensate is an excellent starting point. 
 We also present results for attractive interactions.  Here 
 we find again very good agreement with DMC for the energy of the polaron but less good agreement for the mass.  We attribute this discrepancy to the existence of many-particle bound states
 in the attractive regime \cite{Grusdt2017b,Ardila2015}.
 
\section{Model and proper definition of polaron mass}

Our starting point is a single impurity atom coupled to 
$N$ identical bosons in one dimension, described 
by the Hamiltonian
 \begin{align} 
 \label{Hamiltonian}
     \hat{\mathcal{H}} &= \int \mathrm{d}x\, \odphi \bigg(-\frac{1}{2m}\partial_x^2 +\frac{\gbb}{2}\odphi\ophi-\mu\nonumber \\
     &+\gib \delta(x-\hat{X}) \bigg)\ophi +\frac{\hat{P}^2}{2M}.
\end{align}
Here $m$ ($M$) denotes the mass of the bosons
(impurity atom), $\hat{\phi}(x)$ is the Bose
field operator, $\gbb$ ($\gib$) are the boson-boson (boson-impurity) interaction strength, 
$\hat{X}$ ($\hat{P}$) denotes the
position (momentum) operator of the impurity, and 
$\mu$ is the chemical potential of the bose gas.  
Throughout the paper, we set $\hbar = 1$ and employ periodic boundary conditions of length $L$.
The relative interaction strength is denoted by $\eta = \gib/\gbb$ and we introduce the healing length $\xi = 1/\sqrt{2m\mu}$ and the speed of sound $c = \sqrt{\mu/ m}$.
Expanding the bosonic field operator in Eq.\ (\ref{Hamiltonian}) around a homogenous condensate
as $\ophi = \sqrt{n_0}+ \hat{\xi}(x)$ with $n_0 = N/L$ leads to
the extended Fr\"ohlich Hamiltonian
\cite{Frohlich1954,Shchadilova2016}. 
In this work, we choose a different starting point and consider the effects of the impurity already at the level of the condensate.

Before delving into the solutions of the mean-field equations, it is important to point out some fundamental differences between the ground state of the effective  Fr\"ohlich and the full Hamiltonian (\ref{Hamiltonian}) for finite momentum. For the Fr\"ohlich 
model it is easy to show that for fixed total momentum, the ground state is indeed the polaronic solution \cite{Grusdt2017b,Lee1953,Mahan2000}. 
The situation is very different for (\ref{Hamiltonian}). Indeed the ground state for finite momentum for this case is the uniformly boosted system. To see this 
we introduce the 
potential
$\hat{\Omega}= \hat{\mathcal{H}} - v\hat{P}_\mathrm{tot},$
 with total momentum $\hat{P}_\mathrm{tot} =  \hat{P}_\mathrm{B}+ \hat{P}$ where $\hat{P}_\mathrm{B} = -i  \int \odphi \partial_x \ophi \,\mathrm{d}x$. 
 It is straightforward to see that finding the 
 constrained ground state of (\ref{Hamiltonian}) (with fixed total momentum) is equivalent to finding the unconstrained ground state of $\hat{\Omega}$ for a given $v$ which
 acts as a Lagrange multiplier. Introducing the unitary transformation $
     \hat{U}_\mathrm{cm} = \exp\big(-iM_\mathrm{tot}\hat{X}_\mathrm{cm}v\big),$
 with $M_\mathrm{tot} = Nm+M$ and $\hat{X}_\mathrm{cm} = \frac{1}{M_\mathrm{tot}}(m\int \!\! \mathrm{d}x\, x\odphi\ophi \, +M\hat{X})$ to boost into the center of mass frame 
 one finds
 $   \hat{\Omega} =  \hat{U}^\dagger_\mathrm{cm} \hat{\mathcal{H}}\hat{U}_\mathrm{cm}-\frac{1}{2}M_\mathrm{tot}v^2.$
With this expression, one can clearly relate eigenstates
of $\hat{\cal H}$ with those of $\hat{\Omega}$.
In particular, the ground state for finite momentum (corresponding to finite $v$) is the \textit{boosted}  ground state
and the effective mass of the polaron 
always equals the total mass.  Such a uniformly boosted system
is precluded in the Fr\"ohlich model.

We proceed as in the case of the Fr\"ohlich model and 
eliminate the impurity position operator from 
(\ref{Hamiltonian})
by a Lee-Low-Pines (LLP) \cite{Lee1953} type transformation $\hat{U}_\mathrm{LLP} = \exp(-i\hat{X}\hat{P}_\mathrm{B} )$. Here, in contrast, the \textit{total} momentum of the bosons $\hat{P}_\mathrm{B} $ enters.  $\hat{U}_\mathrm{LLP}$ transforms to 
a co-moving frame, where the impurity is at the origin and its  momentum is transformed to the conserved total momentum of the system  which can be treated
as a c-number $P$. By eliminating the impurity from the problem by an exact transformation, entanglement between the impurity and the condensate is already included on the mean-field level and we do not have to assume a factorised wave function as for example done in \cite{Bruderer-EPL2008,Blinova2013}. 
 At the same time an impurity-mediated interaction between the bosons  $\sim \int\mathrm{d}x\, \big(P-\hat{P}_\mathrm{B}\big)^2/2M$ emerges in the transformed Hamiltonian.
In order to treat this 
it will prove helpful to introduce a Hubbard-Stratonovich field $\hat{u}$, which  gives
 \begin{eqnarray} \label{HamiltonianSLLP}
     \hat{\mathcal{H}}^\mathrm{S}_\mathrm{LLP} = \int \mathrm{d}x\, \odphi \bigg(-\frac{1}{2\mred}
     \partial_x^2+\frac{\gbb}{2}\odphi\ophi-\mu\nonumber\\ +\gib\delta(x)\bigg)\ophi +\hat{u}\big(P-\hat{P}_\mathrm{B} \big)-\frac{1}{2}M\hat{u}^2,\quad
 \end{eqnarray} 
 where $\hat{u}$ satisfies $M{\hat{u}} =P-\hat{P}_\mathrm{B} $,
and can thus be viewed as the impurity velocity. 
$\mred = (M+m)/Mm$ is the reduced mass and we defined  re-scaled healing length $\bar{\xi}= \sqrt{m/\mred}\xi$ and speed of sound $\bar{c} = \sqrt{m/\mred}c$. 

%
\begin{figure}
\includegraphics[]{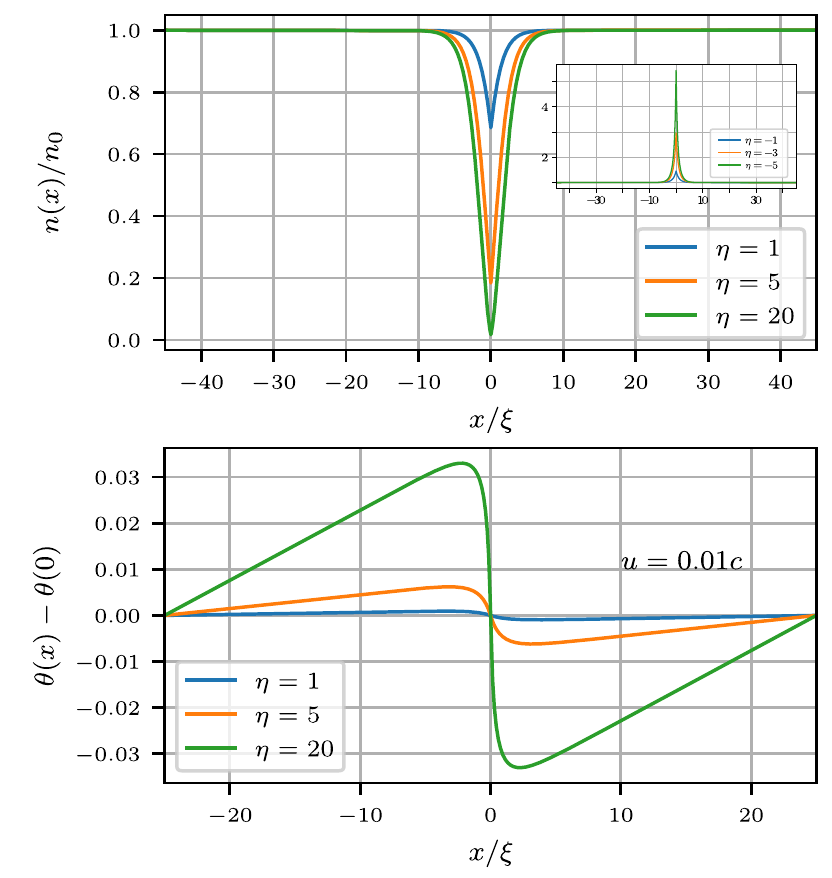}
\caption{Mean-field solution for different interactions and various couplings. All other parameters are as in \cite{Catani2012}, i.e. $M/m = 0.47$, the peak density $n_0 = 7/\mu \mathrm{m}$ and  $\gbb = 2.36 \times10^{-37} \mathrm{Jm}$. For the phase we fixed $u = 0.01c$, which fixes the total momentum on the mean-field level.}\label{Fig:MF-density}
\end{figure}
%

\section{Mean-field solutions}

\subsection{Mean-field equations in the presence of the impurity}

We now expand $\ophi = \phi(x)+ \hat{\xi}(x)$ and $\hat{u} = u + \delta\hat{u}$ where $\phi(x)$ and $u$ are chosen to solve the mean field equations of (\ref{HamiltonianSLLP}), for details see
Appendix A,
\begin{eqnarray}
     &\bigg(-\frac{1}{2\mred}\partial_x^2+\gbb|\phi(x)|^2-\mu +i{u}\partial_x\bigg)\phi(x) = 0&, \label{MF}\\
     &\partial_x\phi(x)\Big{\vert}_{0^-}^{0^+}= 2\mred\gib\phi(0),& \label{jump}
\end{eqnarray}
subject to the boundary conditions
$\phi(\frac{L}{2}) = \phi(-\frac{L}{2})$ and
$|\phi(\pm L/2)|^2 = n_0 +\mathcal{O}(1/L)$. 
Note that in order to 
 remedy the problem of the uniformly boosted system being the ground state we require that the polaron is a \emph{local} quantity. Thus the condensate must be stationary far away from the impurity up to $1/L$ corrections. 
 Solutions of the mean field equations exist in the literature where the phase is not periodic \cite{Hakim1997,Tsuzuki1971} and have been applied to the 1D polaron before \cite{Volosniev2017,Mistakidis2019,Panochko2019}.  The non periodic phase corresponds to unphysical sources at the boundary and leads to wrong predictions such as a negative kinematic polaron mass.
We instead find the 
mean-field solution  of the form $\phi(x) = \sqrt{n(x)}e^{i\theta(x)}$ (see Appendix A for more details).
\begin{align}\label{eq:mfdensity}
     n(x) &= \frac{\mu}{\gbb}\Big(1-\beta \, \mathrm{sech}^2\big( \sqrt{\beta/2}(|x|+ x_0)/\bar{\xi}\big)\Big)
\end{align}
with $\beta = 1-\frac{u^2}{\bar{c}^2}+ \mathcal{O}(1/L^2)$ and $\mu = \gbb n^\mathrm{MF}_0-(\partial_x\theta_1)u + \mathcal{O}(1/L^2)$. If we consider the mean-field solution alone we fix $n^\mathrm{MF}_0 = n_0(1+2\sqrt{2\beta}\bar{\xi}/L(1-\tanh(\sqrt{\beta/2} x_0/\bar{\xi}))) +\mathcal{O}(1/L^2)$, where $n_0=N/L$ is the average density of bosons. Upon considering quantum fluctuations later on, the mean-field density needs to be adjusted. For the phase we find $\theta(x) = \theta_0(x)+(2f(0)-2f(L/2))x/L$ with
\begin{align}
     f(x) = \arctan\Big(\frac{\sqrt{4 u^2 \beta/\bar{c}^2}}{e^{\sqrt{ 2\beta}(x+x_0)/\bar{\xi}}-2\beta +1 }\Big)\nonumber
\end{align} 
for $x>0$ and $\theta_0(x)= 2f(0)-f(-x)$ for  $x <0$.
Finally, we determine $x_0$ through the jump condition for the derivative. 
In the limit $u = 0$ we find for $\gib >0$: $x_0 = \frac{\bar{\xi}}{\sqrt{2}}\log(y)$, with $y = \sqrt{1+8\frac{n_0^2 \bar{\xi}^2}{\eta^2}}+2\frac{\sqrt{2}n_0\bar{\xi}}{\eta}$ and for $\gib <0$, we have $x_0\to x^\mathrm{a}_0 = x_0+i\pi/2\bar{\xi} (2/\beta)^{1/2}$. It is instructive to inserting $x^\mathrm{a}_0$ into (\ref{eq:mfdensity}) and obtain the density profile for the attractive side explicitly
\begin{align}
     n^a(x) &= \frac{\mu}{\gbb}\Big(1+\beta\,\mathrm{csch}^2\big( \sqrt{\beta/2}(|x|+ x_0)/\bar{\xi}\big)\Big).
\end{align}
It becomes apparent that the density far away of the impurity now is lowered instead of increased, and is given by $n^\mathrm{MFa}_0 = n_0(1-2            \sqrt{2\beta}\bar{\xi}/L(\coth(\sqrt{\beta/2} x_0/\bar{\xi})-1)) +\mathcal{O}(1/L^2)$. This seemingly small correction can have a profound impact for $|\eta| \gg 1$. In this limiting case $(\coth(\sqrt{\beta/2} x_0/\bar{\xi})$ divergences and a macroscopic large amount of the bosons aggregates around the impurity. For a finite system this signals a collapse of the condensate onto the impurity. Due to those effect we restrict our analysis of the attractive side to moderate values of $|\gib|$.

In Fig. \ref{Fig:MF-density} mean-field predictions for condensate density and phase are shown for different interaction strength and a slowly moving impurity.
From the analytical solution we can derive
a parameter characterizing the relative condensate deformation 
\begin{equation} \label{eq:deformation_parameter}
   {\eta}/{n_0 \overline{\xi}}=\eta \sqrt{2\overline\gamma} 
\end{equation}
with $\overline{\gamma} = \gamma m_r/m$, where $\gamma=1/(2 n_0^2 \xi^2)$ is the so-called Tonks parameter of the 1D Bose gas \cite{Girardeau1960,Lieb1963a}, which should be less than unity for the Bogoliubov approximation to hold.
The deformation becomes
sizable if  $\eta/n_0 \overline{\xi} \sim 1$.

With the analytical expressions for the condensate density and phase  we can calculate the polaron
energy $E_p=E(g_\textrm{IB})-E(g_\textrm{IB}=0)$
and the effective mass $m^*$ of the polaron
using $M/m^* = \lim_{p\rightarrow 0} (1-\frac{P_\mathrm{B}}{p})$, with $P_\mathrm{B}$ being the mean-field momentum of the condensate, see
\cite{Grusdt2015}.
This gives 
\begin{equation}\label{energy}
    E^\mathrm{r,a}_\mathrm{p} = \gib n_0 \bigg( \frac{|y|\mp 1}{|y|\pm 1}\bigg)^2 + \frac{8}{3}n_0\bar{c}\bigg(\frac{3|y|\pm 1}{(|y|\pm 1)^3} \bigg).
\end{equation}
for the energies of the repulsive ($E_p^{(r)}$, upper sign) and attractive ($E_p^{(a)}$, lower sign) polaron, 
and for the mass:
\begin{equation}\label{meffMF}
    \frac{M}{m^*} = \frac{M(y^2-1) }{8n_0\bar{\xi}\mred \sqrt{2}+M(y^2-1) }.
\end{equation}
%
%
These expressions agree with previous findings in \cite{Volosniev2017} and in \cite{Panochko2019}.
It is interesting to note that for $\eta \rightarrow \infty$, (\ref{energy}) approaches the energy of a dark soliton and the effective mass $m^*$ goes to infinity which is in contrast to results from the extended Fr\"ohlich Hamiltonian \cite{Grusdt2017b}. 
At this point we note that on the attractive side the solution will collapse to a multi particle bound state for $\eta \gg 1$, which can be easily seen by noting $E^\mathrm{a}_\mathrm{p} \rightarrow - \mathcal{1}$ for $\eta \rightarrow - \mathcal{1}$.
\\

\subsection{Boundary conditions}

\begin{figure}[t!]
	\centering
	\includegraphics[width=0.45\textwidth]{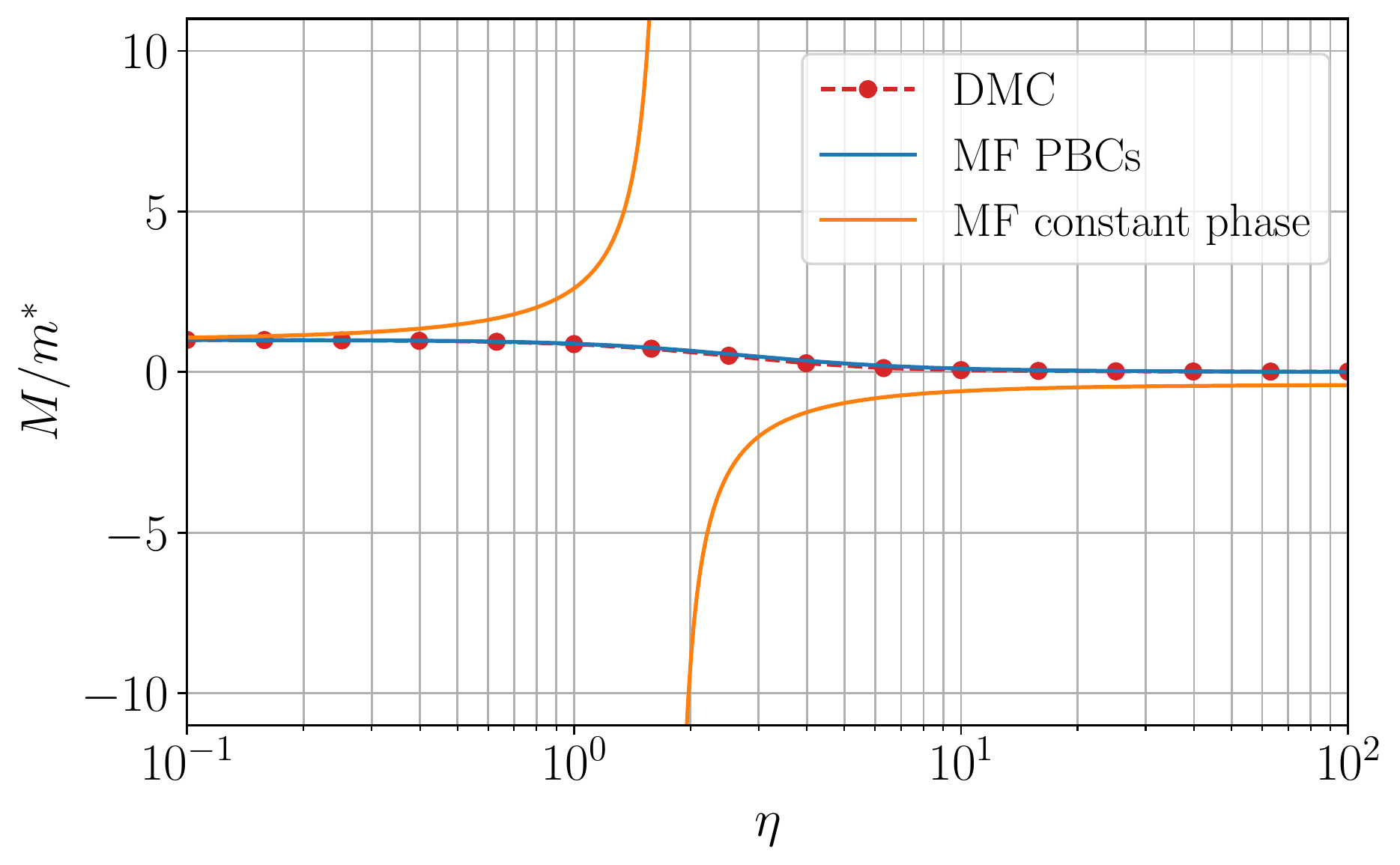}
	\caption{Polaron mass calculated in mean-field approximation with periodic boundary conditions (PBCs) or constant phase far away from the impurity compared to DMC calculated in \cite{Grusdt2017b}. While the solution using periodic boundary conditions agrees very well with the DMC results, the constant phase solution (or non-periodic boundary conditions) yields a non-sensical result. }
	\label{fig:wrong_mass}
\end{figure}

We are now going to address the aforementioned importance of the periodic boundary conditions for correctly calculating the effective mass. When imposing periodic boundary conditions one finds unsurprisingly a constant density far away from the impurity, but in contrast the phase is linearly changing at the order of $1/L$ and therefore not constant. One might be tempted to use a solution where both density and phase are truly constant far away from the impurity (up to exponentially small corrections). A solution with this different boundary condition would still be given by \eqref{PBCtransformation} and \eqref{MFsolution}, but with $\theta_1(x) = 0$. The effective mass can then be deduced from the wave function in the same fashion as was done for periodic boundary conditions and is plotted in Fig.\ \ref{fig:wrong_mass}. Calculating the effective mass in this manner one finds that the effective mass decreases for increasing $\eta$ and it can even become negative. 
This unphysical result is in clear disagreement with DMC results. Besides that, it also contains a phase jump at infinity which introduces a source term there, which is nonsensical.
Addressing this issue from a more technical point of view it becomes apparent that strictly speaking, functional derivatives cannot be taken for the constant-phase solutions. On a mean-field level this can be alleviated by modifying the functional derivatives by exactly this source term as has been done in the context of solitons \cite{Ivan1993,Barashenkov1993}. Another possible way to deal with the phase issue is to integrate the phase out as has been done in \cite{Panochko2019}.  Upon considering quantum fluctuations on top of the mean-field solution none of the above mentioned methods allow a straightforward generalization.  We found expanding about a periodic mean-field solution to be indispensable for the Bogoliubov theory.

We note that this issue persists when
extracting the mass from the total momentum dependence of the mean-field energy of the system.  That is, when enforcing the non-periodic phase, and expanding the total mean-field energy to quadratic order in the total momentum as $E \approx E_0+ \frac{p^2}{2m^*}$, one obtains an incorrect result for the polaron mass 
$m^*$.    On the other hand, when extracting the polaron mass from expanding in $u$ as
$E \approx E_0 + \frac{1}{2} m^* u^2$
one fortuitously obtains the correct result with both periodic and non-periodic \cite{Mistakidis2019} mean-field 
solutions.

These difficulties can be traced to the fact that without the phase correction, the mean-field equations of motion do not form a Hamiltonian system.  For
the full quantum system, one can deduce that the
fundamental relation 
\begin{align}
\label{ham}
\frac{dE}{dp}=u 
\end{align}
holds exactly by
the Feynman-Hellman theorem.  Incidentally, this
relation can be used to obtain 
$M/m^* = \lim_{p\rightarrow 0} (1-\frac{P_\mathrm{B}}{p})$ which is used routinely to compute the polaron mass.
With periodic boundary conditions, one
retains  the exact relation (\ref{ham}) within mean-field theory.
On the other hand, when the non-periodic solution is used 
a short calculation gives the relation:
\begin{align}
\frac{dE_{\rm np}}{dp} = u - u \bar{n} 
\frac{d}{dp} \Delta \theta  
\end{align}
where $E_{\rm np}$ is the total mean-field energy of the non-periodic state,
$\bar{n}$ is the average density, and $\Delta \theta$ is the phase change across the condensate.\\

\section{Quantum fluctuations}

\begin{figure}
\includegraphics{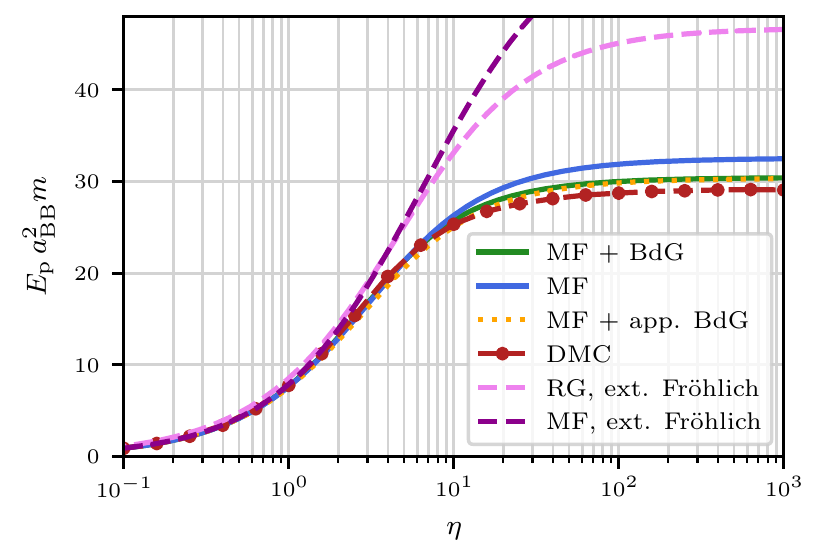}
\includegraphics{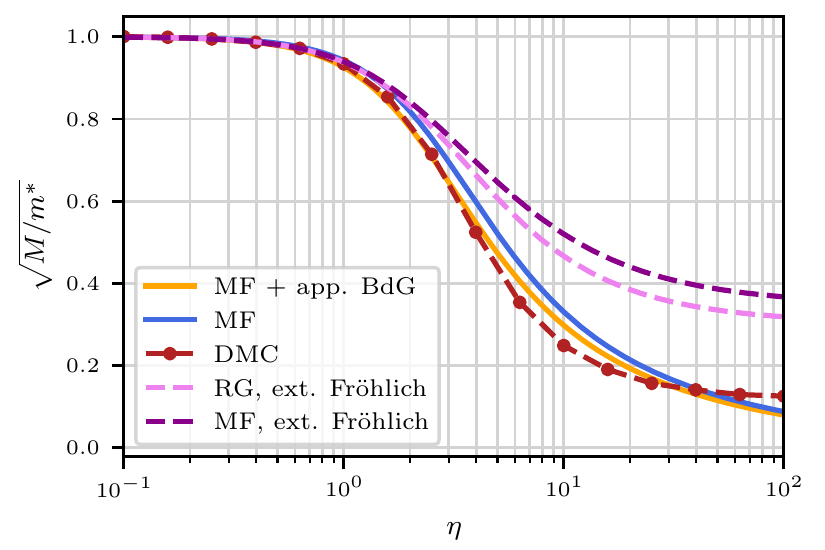}
\caption{Polaron energy (top) and effective mass (bottom) for the repulsive polaron. The curves are obtained using different theoretical methods all parameters are as in \cite{Catani2012}, where $\gamma\approx 0.438$. The DMC, RG and MF (both based on the extended Fr\"ohlich model) curves were calculated in \cite{Grusdt2017b}. We find exceptional agreement with the DMC results for the energy as well as the effective mass when expanding around the right mean-field solutions and including quantum fluctuations. Only for the very strong coupling regime we do not predict a saturation of the effective mass. The condensate deformation becomes relevant for $\eta / n_0 \overline{\xi}>1$ \eqref{eq:deformation_parameter}, corresponding here to $\eta > 1.9$, where predictions from the ext.~Fr\"ohlich model start to deviate from the full model.\label{Erepfig}}
\end{figure}

After expanding the fields 
in $\hat{\cal H}^\textrm{S}_\textrm{LLP}$ in the quantum fluctuations 
we find up to second order in $\odxi$ and $\delta\hat{u}$
\begin{eqnarray} \label{expandedSLLP}
     &&\hat{\mathcal{H}}^\mathrm{S}_\mathrm{LLP} = \int\!\! \mathrm{d}x\, \bigg[\odxi \bigg(-\frac{1}{2\mred}\partial_x^2+2\gbb|\phi(x)|^2-\mu+
     \nonumber\\
     &&+\gib\delta(x) + i u\partial_x\bigg)\oxi + \frac{\gbb}{2}\Big(\phi(x)^2\odxi^2+h.a.
     \Big)\bigg]\nonumber\\
     &&{-i\delta\hat{u}}\int\!\!\mathrm{d}x\Big(\odxi \partial_x\phi(x) +\phi^*(x) \partial_x\oxi  \Big)-\frac{1}{2}M\delta\hat{u}^2,
\end{eqnarray}
with $M\delta \hat{u} = -i \int \big[\phi^*(x) \partial_x\oxi +\odxi\partial_x\phi(x)\big] \, \mathrm{d}x + \mathcal{O}(\oxi^2)$, which can be diagonalized by a Bogoliubov rotation to a generalized basis of \emph{phonons on a deformed background}. We note that for distances far away from the impurity i.e.~$\vert x\vert \rightarrow \infty$ these phonons look like the ones of a homogeneous BEC. This allows us to extract the quantum depletion (see \cite{Salasnich2016} for a detailed discussion on how to regularise the arising UV-divergences of the zero point energy). We find for the quantum-corrected density far away from the impurity $n_0 = n^\mathrm{MF}_0 +\frac{1}{\pi} \sqrt{\mred\gbb n^\mathrm{MF}_0 }$ and thus we have to adjust the mean-field density accordingly. To diagonalize (\ref{expandedSLLP}) we note
 that all terms involving $\delta\hat{u}$ become non local and thus difficult to handle in general, except for the special case $p=0$. This enables us to diagonalize (\ref{expandedSLLP}) and to calculate the polaron energy for $p=0$. For a moving impurity 
we introduce an approximation setting $\delta\hat{u}=0$ 
and keep $u$ as variational parameter in the mean-field equations. After diagonalizing the remaining quadratic Hamiltonian (\ref{expandedSLLP}) $u$ is determined self-consistently:
\begin{eqnarray}
Mu &=& P - \langle \hat{P}_\mathrm{B} \rangle,\\
\langle \hat{P}_\mathrm{B} \rangle  &=& -i \int \phi^*(x) \partial_x\phi(x) \,\mathrm{d}x -i \langle\int \odxi \partial_x\oxi \,\mathrm{d}x\rangle,
\nonumber
\end{eqnarray}
where the expectation value is taken with respect to the phonon vacuum. For a more detailed description we refer to Appendix B. Then it is straightforward to calculate the effective mass including the quantum corrections $M/m^* = Mu/p$.
As can be seen in Fig.\ref{Erepfig} a), where the energies of the full and approximate solution of the BdG equations are shown,
the approximate treatment of the Hubbard-Stratonovich field is very good.\\

%
%
\begin{figure}[htb]
\includegraphics{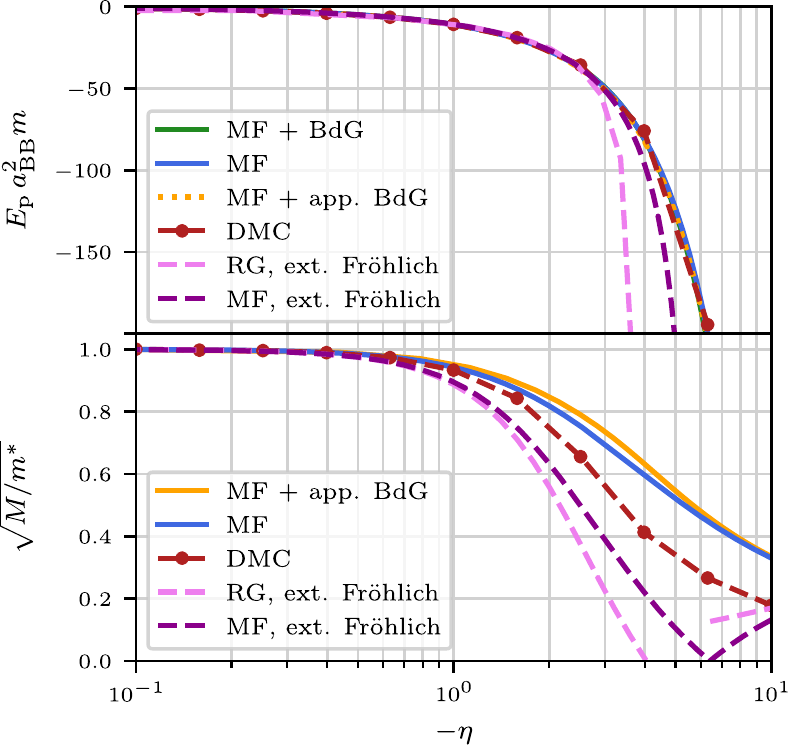}
\caption{Polaron energy (top) and mass (bottom) calculated using different methods for attractive impurity couplings. All parameters are as in \cite{Catani2012} and the RG, MF (based on the extended Fr\"ohlich model) and DMC curves were calculated in \cite{Grusdt2017b}. For the polaron energy  a surprisingly good agreement is achieved with the DMC results, while the agreemenat is less good for the mass.  We explain this by the collapse of the solution to a multi-particle bound state, where we do not expect the mean-field solution to be a good approximation. Futhermore we do not observe the transition from the attractive to the repulsive polaron observed in the Fr\"ohlich model, signaled by the breakdown of the RG treatment. For more detail on this transition we refer to \cite{Grusdt2017b,Shchadilova2016}.
}
\end{figure}
%

\section{Discussion \& Summary}

Figs. \ref{Erepfig} show that already the mean-field solution improves the agreement with DMC simulations significantly 
for $g_\textrm{IB}>0$  as compared to the Fr\"ohlich model. Including quantum fluctuations leads to almost perfect
agreement for the energy. We find however
that the effective mass diverges for $g_\textrm{IB}\to\infty$, even after including quantum fluctuations, which seem to be in contrast to the DMC results \cite{Grusdt2017b}. This divergence is a characteristic of the 1D geometry and is, for example, also observed in the Tonks-limit \cite{Parisi2017}. One would naively expect this to happen since for $\eta \gg 1$  the condensate is split into two halves by the impurity, preventing any transport of the condensate across it. The only possible contribution could come from tunnelling which is highly suppressed for $\eta \gg 1$. The same reasoning explains why the quantum correction to the effective mass is  most significant for intermediate couplings since here the classical current is reduced by the strong condensate deformation, but tunnelling is still relevant. The question whether the effective mass actually saturates  remains open and other approaches such as DMRG could shed more light on this. Note that these arguments rely on treating the system as one-dimensional.  For experimental systems in the one-dimensional regime, we expect that transverse modes may
become important for the limiting behaviour of $M/m^*$. This analysis has to be done on a case to case basis and we want to stress that all our calculations are bench marked against strict 1D numerical quantum Monte-Carlo results. For a detailed discussion on the influence of the transverse mode and when it's admissible to treat the system investigated in \cite{Catani2012} as strictly one dimensional we refer to the detailed discussion in \cite{Grusdt2017b}.
Another quantity of experimental relevance 
\cite{Catani2012}
is the axial width
of the polaron
$\left(\langle \hat{X}^2 \rangle - \langle \hat{X} \rangle^2 \right)^{1/2}$.  In the present work, which is carried out
in the LLP frame and requires translational invariance, such a quantity is infinite.     Including a trap potential for the impurity is beyond the scope of the present work, but could be  addressed by using a variational ansatz that is a superposition
of ground states (of the infinite system) with different total momenta.
On the other hand,
studies that do not invoke the LLP transformation can lead to symmetry-broken mean-field states with finite values of the axial width \cite{Cucchietti2006,Blinova2013} even without
a trap, but these neglect impurity-BEC entanglement.

In summary, we have shown that a non-perturbative description of the Bose polaron in 1D requires taking  into account the backaction to the condensate while keeping the impurity-BEC entanglement. Since the density of phonons defined on such a deformed background remains small, their intrinsic interactions can be neglected to good approximation. Our approach provides a quantitatively accurate and, to a large extent, analytical description of Bose polarons even for strong impurity-boson interactions as long as the boson-boson interactions remain weak. Those findings suggest that a similar method could be used to gain more insight into the polaron formation following a sudden quench. We expect that it will also allow a good description in 3D at and beyond the critical strength of the impurity-boson interaction for self-trapping.

\section*{Acknowledgments}

We would like to thank G. Astrakharchik and J. Brand for useful discussions and insights. We also thank G. Astrakharchik and E. Demler for providing the DMC and RG data of Ref.\ \cite{Grusdt2017b}. M.W. and M.F. acknowledge financial support by the DFG through the SFB/TR 185, project number 277625399.
J.J.\ is grateful  for support from EPSRC under Grant
EP/R513052/1.  R.B. is grateful for support from a Cecilia Tanner
Research Impulse Grant  and the Aspen Center for Physics,  which is
supported by NSF PHY1607611 and the Simons Foundation.

\subsection{Appendix A}
In this Appendix, we provide some details on the solution of the mean-field equations used in the main text. The mean-field equations that need to be solved are
\begin{align}\label{MFeqn}
     \bigg(-\frac{1}{2m_\mathrm{red}} \partial_x^2 &+\gbb|\phi(x)|^2-\mu + iu \partial_x \bigg)\phi(x) = 0 \nonumber\\
     \partial_x \phi(x)\Big{\vert}_{0-}^{0+}&= 2m_\mathrm{red}\gib \phi(0) \nonumber\\
     \phi(L/2) &= \phi(-L/2)\nonumber\\
     \quad Mu &= p - P_\mathrm{B}.
\end{align}
If we do not require periodic boundary conditions analytical solutions of the form $|\phi(x)|e^{i\theta_0(x)}$ can be found in the literature \cite{Hakim1997,Tsuzuki1971}. To make use of those solutions we make the following ansatz
\begin{equation}\label{PBCtransformation}
    \phi(x) = \tilde{\phi}(x)e^{i\theta_1(x)},
\end{equation}
where we introduced $\theta_1(x)$ which will be of $\mathcal{O}(x/L)$ and is fixed later on to ensure periodicity of the phase for the mean-field solutions, giving the overall phase $\theta(x) = \theta_0(x)+\theta_1(x)$. Upon inserting our ansatz into (\ref{MFeqn}) we arrive at
\begin{align}\label{MFeqnphase}
     \bigg(-\frac{1}{2m_\mathrm{red}} \partial_x^2
     &+\gbb|\tilde{\phi}(x)|^2-\tilde{\mu}+ i\tilde{u}
    \partial_x \bigg)\tilde{\phi}(x) = 0 \nonumber\\
    \partial_x \tilde{\phi}(x)\Big{\vert}_{0-}^{0+}&= 2m_\mathrm{red}\gib  \tilde{\phi}(0) \nonumber \\ e^{i\theta_1(L)}\tilde{\phi}(L/2) &= \tilde{\phi}(-L/2)
\end{align}
with the re-definitions $\tilde{\mu} = \mu +(\partial_x\theta_1) u/M+\mathcal{O}(1/L^2) $ and $\tilde{u} = u-(\partial_x\theta_1)/(\mred) $. The solution for this problem is now given by \cite{Hakim1997,Tsuzuki1971}
\begin{align}\label{MFsolution}
        |\tilde{\phi}(x)| &=\sqrt{\mu/\gbb}\Big(1-\beta \, \mathrm{sech}^2\big(\nonumber \sqrt{\beta/2}(|x|+ x_0)/\bar{\xi}\big)\Big)^{1/2},\nonumber\\
    \theta_0(x) &= 
    \begin{cases}
        f(x)        \quad &x>0 \\
        2f(0)-f(-x) \quad &x<0,
    \end{cases} \notag \\
     f(x) &=  \arctan\big(\frac{\sqrt{4 u^2 \beta/\bar{c}^2}}{e^{\sqrt{ 2\beta}(x+x_0)/\bar{\xi}}-2\beta +1 }\big) 
\end{align}
with $\beta = 1-\frac{u^2}{\bar{c}^2}+ \mathcal{O}(1/L^2)$ and $\mu = \gbb n^\mathrm{MF}_0-(\partial_x\theta_1(x)) \,u/M + \mathcal{O}(1/L^2)$. 
The jump condition determines $x_0$ through a polynomial of order three, but only one solution is stable. It is possible to extract quantities like the critical momentum from here, for a detailed discussion of this we refer to \cite{Hakim1997}. For finite momentum, the condition for $x_0$ has to be solved numerically but in the limit $p \rightarrow 0$ we can find the analytical solutions stated in the paper.
If we consider the mean-field solution alone and require the number of condensed particles $N$ to stay constant on the 
mean-field level we fix $n^\mathrm{MF}_0 = n_0 \left[1+2\sqrt{2\beta}\bar{\xi}/L \left(1-\tanh(\sqrt{\beta/2} x_0/\bar{\xi})\right) \right] +\mathcal{O}(1/L^2)$. Lastly, we fix $\theta_1(x)$ to ensure the periodicity of the phase by 
\begin{eqnarray}
         \theta_1(x) =2 \left[f(0)-f(L/2)\right]  \frac{x}{L}.
\end{eqnarray}
At this point we note that the $1/L$ corrections are indeed important when calculating physical quantities. This can be seen by considering the Boson momentum $P_\mathrm{B}= \int n(x)\partial_x \theta(x)\,\mathrm{d}x  = \int n(x)\partial_x \theta_0(x)\,\mathrm{d}x + n_0 \left[2(f(0)-f(L/2)\right]$. Form there we can derive the expressions for $m^*$ and $E_p$ given in the main text, which are both defined in the limit $p \rightarrow 0$, which allows us to state them fully analytically.  

\subsection{Appendix B}
In the following, we give a short overview of the methods used to obtain the quantum corrections to the mean-field solutions. The major steps have been outlined in the main text, and thus we focus on the numerical details. An extensive overview of the techniques used here can be found in \cite{Xiao2009}. We note that this is equivalent to solving the resulting Bogolibouv-de Gennes equations.
We start by discretizing $\hat{\mathcal{H}}^\mathrm{S}_\mathrm{LLP}$ from the main text after either making the approximation of treating $u$ as a variational parameter or for $p=0$ integrating out the $\hat{u}$-field. For all numerical results presented here, the discretization was
done in real space and is therefore 
straightforward apart from the delta distribution, which was approximated by a Kronecker delta in the following way $\delta(x) \rightarrow \delta_{i,0}/a$, where $a$ is the discretization. This comes at the expense of not accounting correctly for the UV behavior. The deviation from the continuum UV behavior is due to discretizing the derivative operators. Nevertheless, for the observables we are interested in here the UV behavior is not essential, and we found fast convergence; thus, the diagonalization in real space is justified. For notational convenience, we omit the hats on all discretized operators. After discretization the Hamiltonian can be written as
\begin{align}
    {{H}}^\mathrm{S}_\mathrm{LLP} &=\sum_{ij}\bigg[ A_{ij}\phi_i^\dagger\phi_j + \frac{1}{2}(B_{ij}\phi_i^\dagger\phi^\dagger_j + B^*_{ij}\phi_i\phi_j )\bigg] \notag \\
    &= \frac{1}{2}\mathbf{\Phi}^\dagger M \mathbf{\Phi} -\frac{1}{2}\mathrm{Tr}(A),
\end{align}
where $ \mathbf{\Phi}^\dagger= [\phi_1^\dagger,\phi_2^\dagger,...\phi_n^\dagger,\phi_1,\phi_2,...\phi_n]$ is the discrete version of $\hat{\xi}(x)$ and $M$ is the semi positive definite matrix 
\begin{align}
       M =  \begin{bmatrix} 
A & B \\
B^* & A^* 
\end{bmatrix}. 
\end{align}
At this point we already note that the trace term is of fundamental importance in 1D since it renders results like the zero point energy finite without performing additional regularization. Following the steps outlined in \cite{Xiao2009} we now diagonalize 
\begin{align}
\nu M =  
\begin{bmatrix} 
A & B \\
-B^* & -A^* 
\end{bmatrix}, 
\end{align}
and thus find $T$ such that $T^\dagger M T = \mathrm{diag}(\omega_1,\omega_2,...,\omega_n,\omega_1,\omega_2,...,\omega_n)$, while guaranteeing $T^\dagger\nu T = \nu$, which allows us to introduce new bosonic operators $\mathbf{\Psi}^\dagger= [b_1^\dagger,b_2^\dagger,...b_n^\dagger,b_1,b_2,...b_n]$ through
\begin{equation}
    \mathbf{\Phi} = T \mathbf{\Psi},
\end{equation}
for which the Hamiltonian takes diagonal form. The new operators $b_i$ can be interpreted as quasiparticle-like bosonic excitations with eigenenergy $\omega_i$. For a stable polaron the energy of those excitations is minimized, i.e.\ the system is in its vacuum state $|0\rangle$ with respect to the $b_i$. From here it is then easy to verify that the quantum corrections to the expectation value of an observable of the form ${O}_Q= \sum_{ij} O_{ij}\phi^\dagger_i\phi_j$ is 
\begin{align}
    \langle {O_Q} \rangle &= \langle0| \mathbf{\Psi}^\dagger T^\dagger \begin{bmatrix} 
O & 0 \\
0 & 0
\end{bmatrix} T\mathbf{\Psi} |0\rangle \notag \\
&= \langle0| \mathbf{\Psi}^\dagger \begin{bmatrix} 
C& D \\
E & F
\end{bmatrix} \mathbf{\Psi} |0\rangle \notag\\
&= \mathrm{Tr}(F).
\end{align}
To conclude this section, we will comment on the IR (infrared) divergences, that are characteristic in 1D systems and how they are dealt with here. First, we note that quantities like the two-point function
\begin{equation}
    \langle \phi_i^\dagger\phi_i \rangle = \langle0| (\mathbf{\Psi}^\dagger T^\dagger)_i (T\mathbf{\Psi})_i |0\rangle \sim L,
\end{equation}
are indeed IR divergent in our treatment. For the global quantities and $p=0$, this can be dealt with as outlined in the main text by considering the zero point energy
\begin{equation}
    E = \frac{1}{2}(\sum_i \omega_i-\mathrm{Tr}(A)),
\end{equation}
which is UV and IR finite and then taking adequate derivatives (i.e.\ with respect to the chemical potential for the depletion). When considering $\hat{\mathcal{H}}^\mathrm{S}_\mathrm{LLP}$ for $p \neq 0$ without any approximations, the phonon momentum seems to be IR divergent and also for the polaron energy we found a system size dependence. 
Lastly, we remark that in the approximate treatment, i.e.\ when viewing $u$ as a variational parameter, the phonon momentum remains IR and UV finite. Therefore, we conclude that all results presented in the main text are cut-off independent, and no divergences occur. 
\bibliographystyle{apsrev4-1} 
\bibliography{library} %
\end{document}